\documentclass{ws-ijmpd}

\begin{document}

\markboth{J.-P.~Lenain, M.~K.~Daniel, C.~Boisson, P.~M.~Chadwick, H.~Sol \& M.~J.~Ward}
{VHE AGN SSC modeling tour}

%
\catchline{}{}{}{}{}
%

\title{VERY HIGH ENERGY ACTIVE GALACTIC NUCLEI SYNCHROTRON SELF-COMPTON MODELING TOUR}

\author{J.-P.~Lenain\footnote{jean-philippe.lenain@obspm.fr}, C.~Boisson, H.~Sol}
\address{LUTH, Observatoire de Paris, Universit\'e Paris Diderot; 5 Place Jules Janssen\\
92190 Meudon, France}

\author{M.~K.~Daniel, P.~M.~Chadwick, M.~J.~Ward}
\address{Department of Physics and Astronomy, Durham University, South Road\\
Durham DH1 3LE, U.K.}

\maketitle

\begin{history}
\received{Day Month Year}
\revised{Day Month Year}
\comby{}
\end{history}

\begin{abstract}
The current very high energy (VHE; $E>100$\,GeV) experiments have tremendously increased the number of detected extragalactic sources. We present a synchrotron self-Compton modeling tour of the active galactic nuclei currently established as VHE emitters so far, and investigate possible correlations among the intrinsic and derived parameters.
\end{abstract}


Active galactic nuclei are known to emit non-thermal radiation in a wide range of frequencies. In recent years, the current generation of imaging atmospheric \v{C}erenkov telescopes (IACT) probing very high energy (VHE; $E>100$\,GeV) $\gamma$-rays have detected more than twenty extragalactic sources, the major part of which belongs to the blazar class.

We present here a modeling tour of these VHE emitting sources, using a steady-state synchrotron self-Compton (SSC) model consistently, and study possible links between the derived parameters among the sources.

\section{The model}

The high energy emitting zone is modeled as a blob of plasma radiating at all frequencies through SSC process, propagating through an inhomogeneous extended jet responsible for the low frequency radiation. The blob is comprised of a relativistic population of electrons/positrons. A tangled magnetic field fills up the assumed spherical blob. The $\gamma\gamma$ absorption by the extragalactic background light is accounted for and corrected using the low-IR model from Refs.~\refcite{2002A+A...386....1K,2004A+A...413..807K}. For more details on the SSC model, please refer to Ref.~\refcite{2001A+A...367..809K}.

\section{The sources}	

We retrieved VHE data from the literature for 24 AGNs known to emit VHE $\gamma$-rays, observed by MAGIC, VERITAS, CANGAROO, or H.E.S.S. We also gathered archival multi-wavelength data for these sources, using SIMBAD\footnote{http://simbad.u-strasbg.fr/simbad/}, NED\footnote{http://nedwww.ipac.caltech.edu/} and Ref.~\refcite{2009A+A...495..691M}\footnote{http://www.asdc.asi.it/bzcat/}.

\begin{figure}
\centerline{\psfig{file=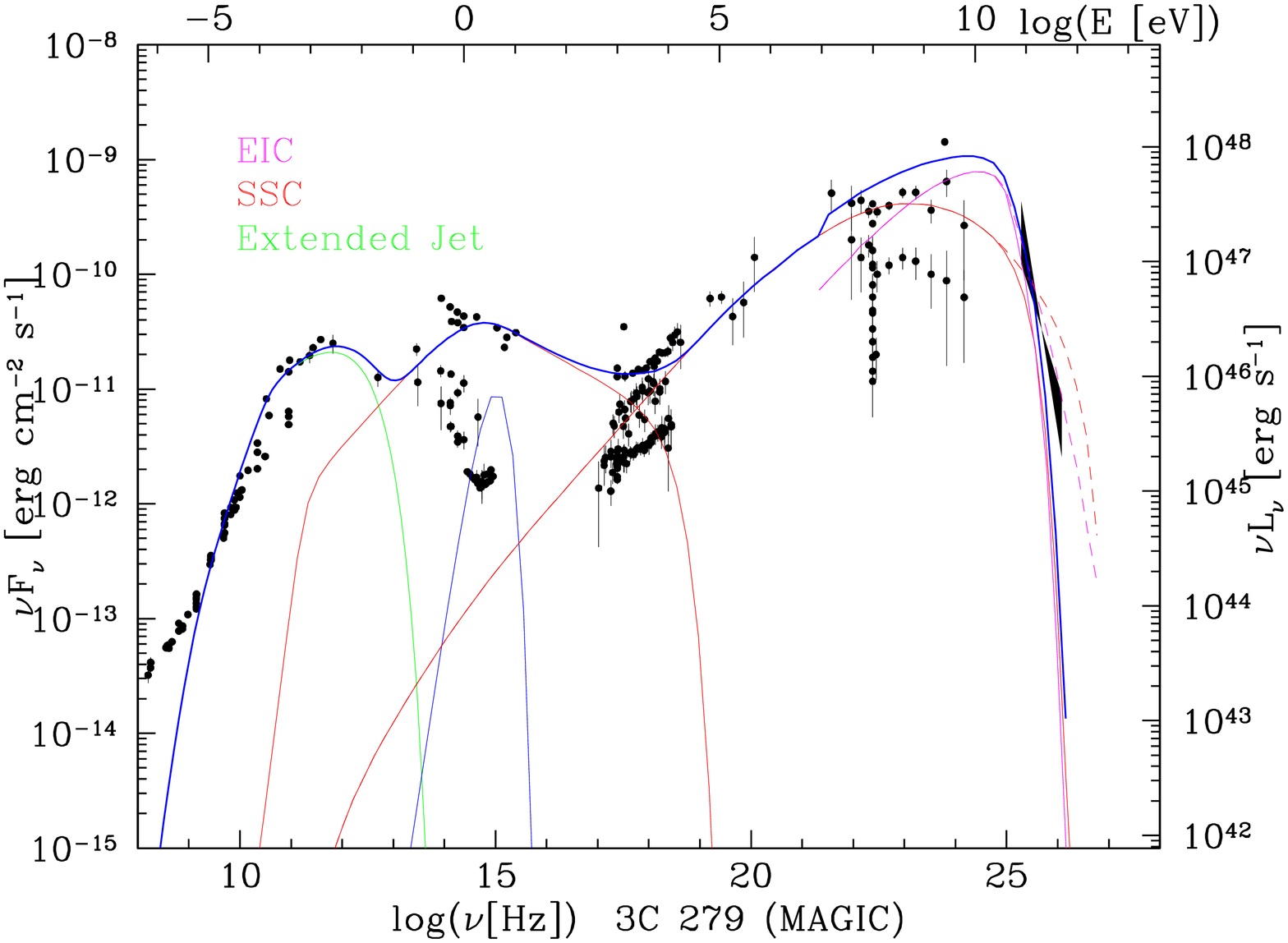,angle=-90,width=0.45\textwidth}
\hspace{0.02\textwidth}
\psfig{file=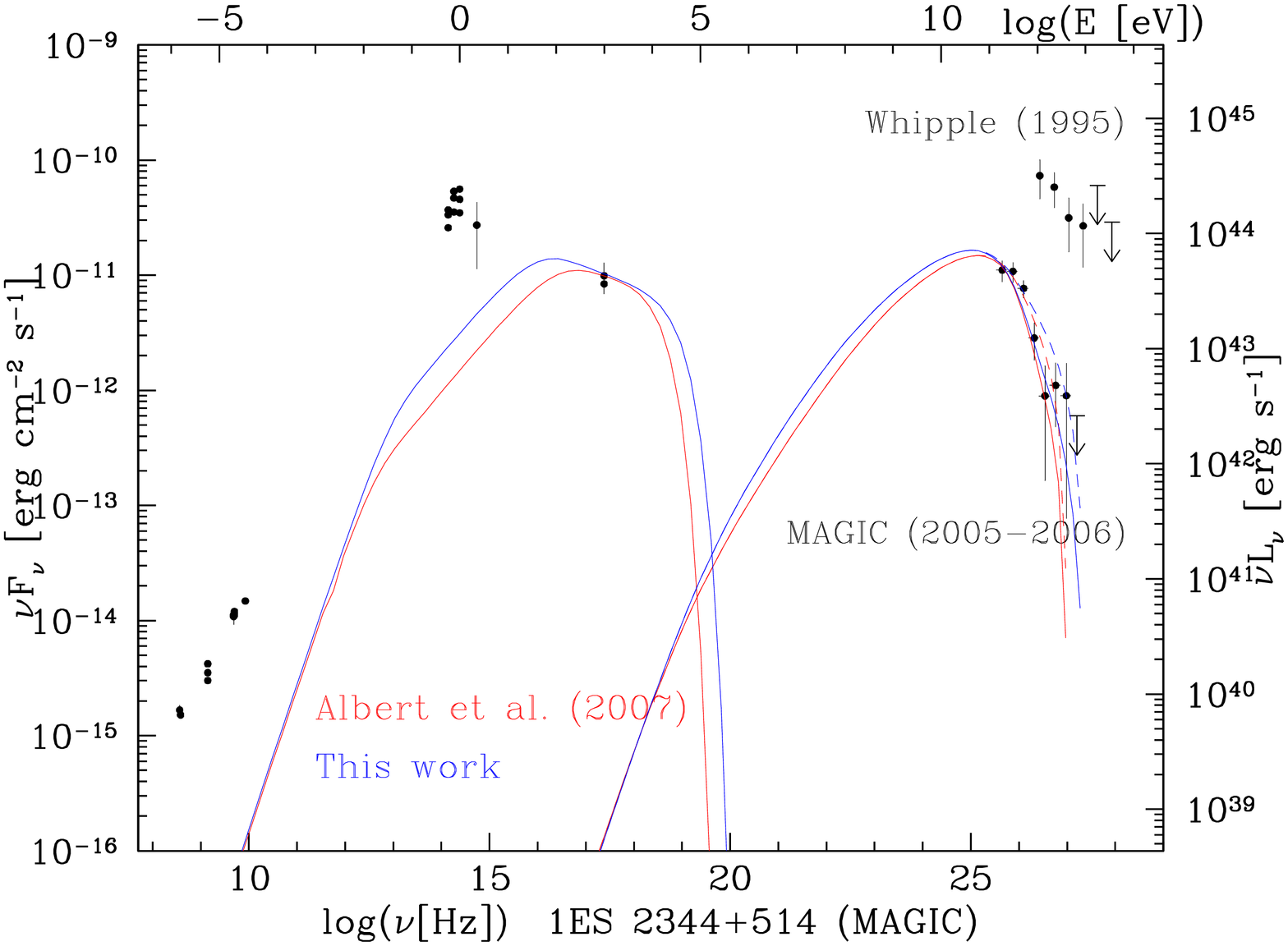,angle=-90,width=0.45\textwidth}}
\vspace*{8pt}
\caption{{\it Left}: SED of 3C\,279. {\it Right}: SED of 1ES\,2344$+$514.
\label{fig-3C279-1ES2344}}
\end{figure}

Consistently, we strictly applied the same SSC model to all these sources. In order to reduce the parameter space available, for all the sources we choose to fix the value of the magnetic field close to $B\sim0.1$\,G and the value of the Doppler factor $\delta_b\sim20$. The first index of the electron energy distribution is $n_1=2.0$, assuming a Fermi\,I acceleration by a strong shock.

\begin{figure}
\centerline{\psfig{file=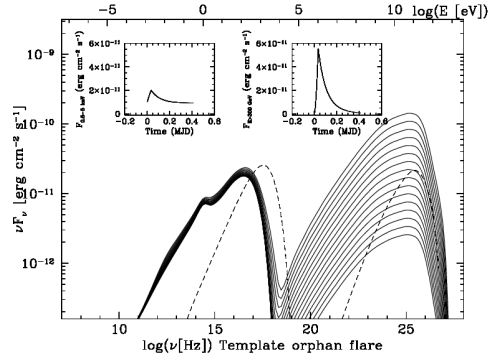,width=0.5\textwidth}}
\vspace*{8pt}
\caption{SED of a template solution for an orphan VHE flare event. The inlays represent the corresponding light curves in X-rays and at VHE. Each solid line shows a snapshot of the evolving SED, with the whole flare lasting for $\sim10$\,h. For comparison, the dashed lines show the quiescent state of PKS\,2155$-$304 as observed with H.E.S.S. in 2003.
\label{fig-OrphanFlare}}
\end{figure}

We show in Fig.~\ref{fig-3C279-1ES2344} an instance of the application of the SSC model to 3C\,279 and 1ES\,2344$+$514. In the case of 3C\,279\cite{2008Sci...320.1752M}, the nature of its high energy component is still uncertain. Leptonic scenarios seem to require extreme parameters if equipartition between the magnetic energy density and the particle energy density is to be sustained\cite{2008AIPC.1085..427B,tmpBoettcher_TheseProc}. We present here a solution, far from the equipartition ($U_B/U_\mathrm{part}\sim10^{-3}$), mixing SSC and external inverse Compton scattering off a blackbody radiation field, which could be the accretion disk, or the broad line region for instance. We also present in Fig.~\ref{fig-3C279-1ES2344} the SED of 1ES\,2344$+$514 as modeled in Ref.~\refcite{2007ApJ...662..892A} and another solution using our SSC model. Compared to Ref.~\refcite{2007ApJ...662..892A}, we changed the Doppler factor $\delta_b=8.4 \rightarrow 20$ and the required size of the emitting region is smaller ($r_b\sim10^{16}$\,cm$\rightarrow 3.5 \times 10^{15}$\,cm).
The point here is to show that solutions are degenerate even from a simple one-zone SSC model as applied here. Long and short term monitoring should allow to solve such degeneracy.

\section{Time dependent modeling for an orphan flare event}

We also applied a time dependent SSC model, which description can be found in Ref.~\refcite{2003A+A...410..101K}, to the well known $\gamma$-ray orphan flare event observed in 1ES\,1959$+$650\cite{2005ApJ...621..181D,2004ApJ...601..151K}. Extremizing the parameters derived using this model to describe the second major VHE flare observed in PKS\,2155$-$304 on July 30, 2006 with H.E.S.S.\cite{2008AIPC.1085..415L}, we found a reasonable template solution describing a VHE orphan burst as seen in 1ES\,1959$+$650 (see Fig.~\ref{fig-OrphanFlare}). We show for comparison the quiescent state of PKS\,2155$-$304 as observed in 2003 with H.E.S.S.\cite{2005A+A...442..895A}

\section{Correlation studies}

Based on this sample of sources, we investigated possible correlations among the intrinsic and derived parameters. As caveats, one has to remind that this sample is biased towards flaring states, due to the strategy of observations of IACTs, and is biased towards high-frequency-peaked blazars (HBL). The parameters derived from this SSC study is also obviously model dependent.

\begin{figure}[h]
\centerline{\psfig{file=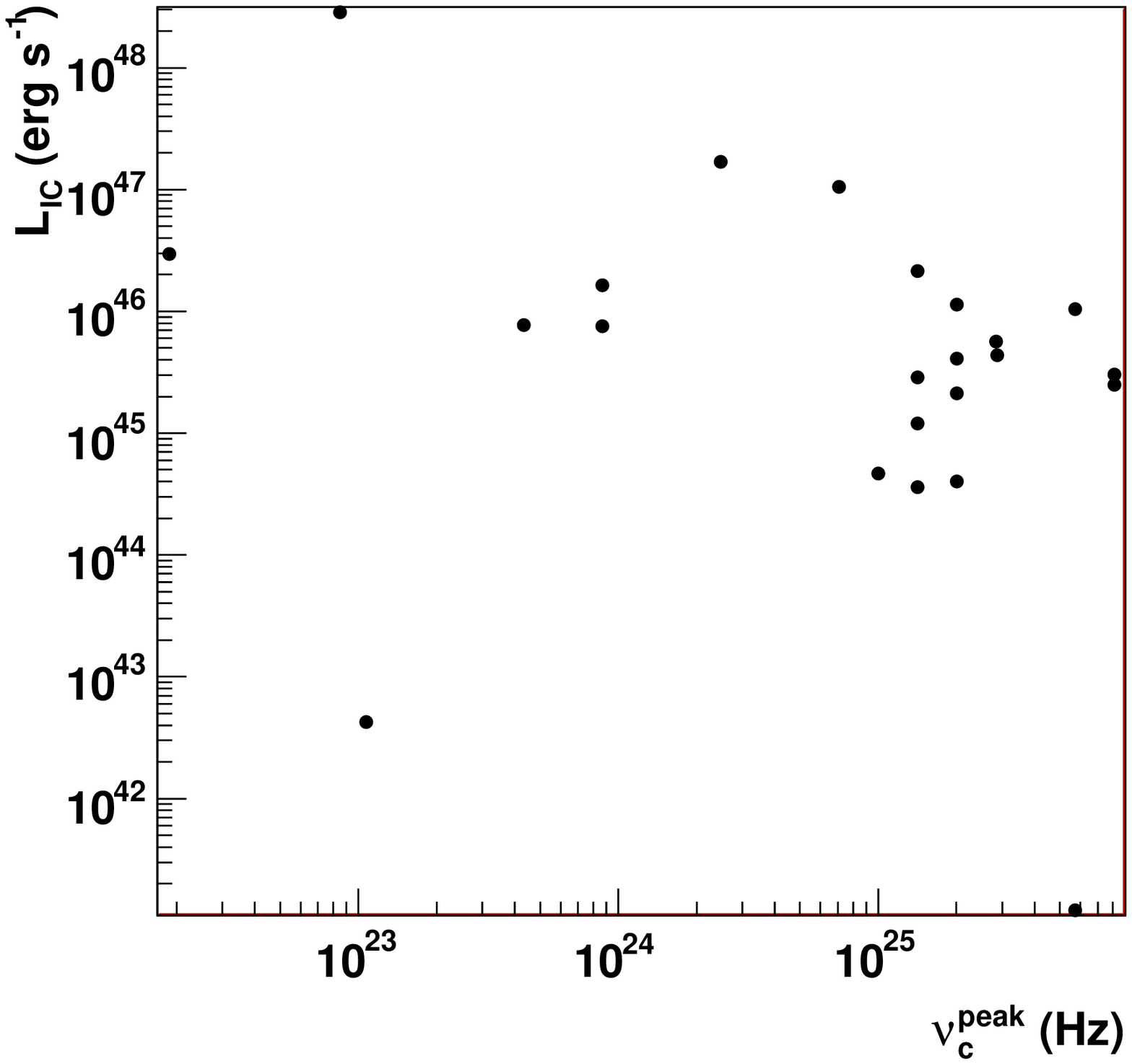,width=0.45\textwidth}
\hspace{0.02\textwidth}
\psfig{file=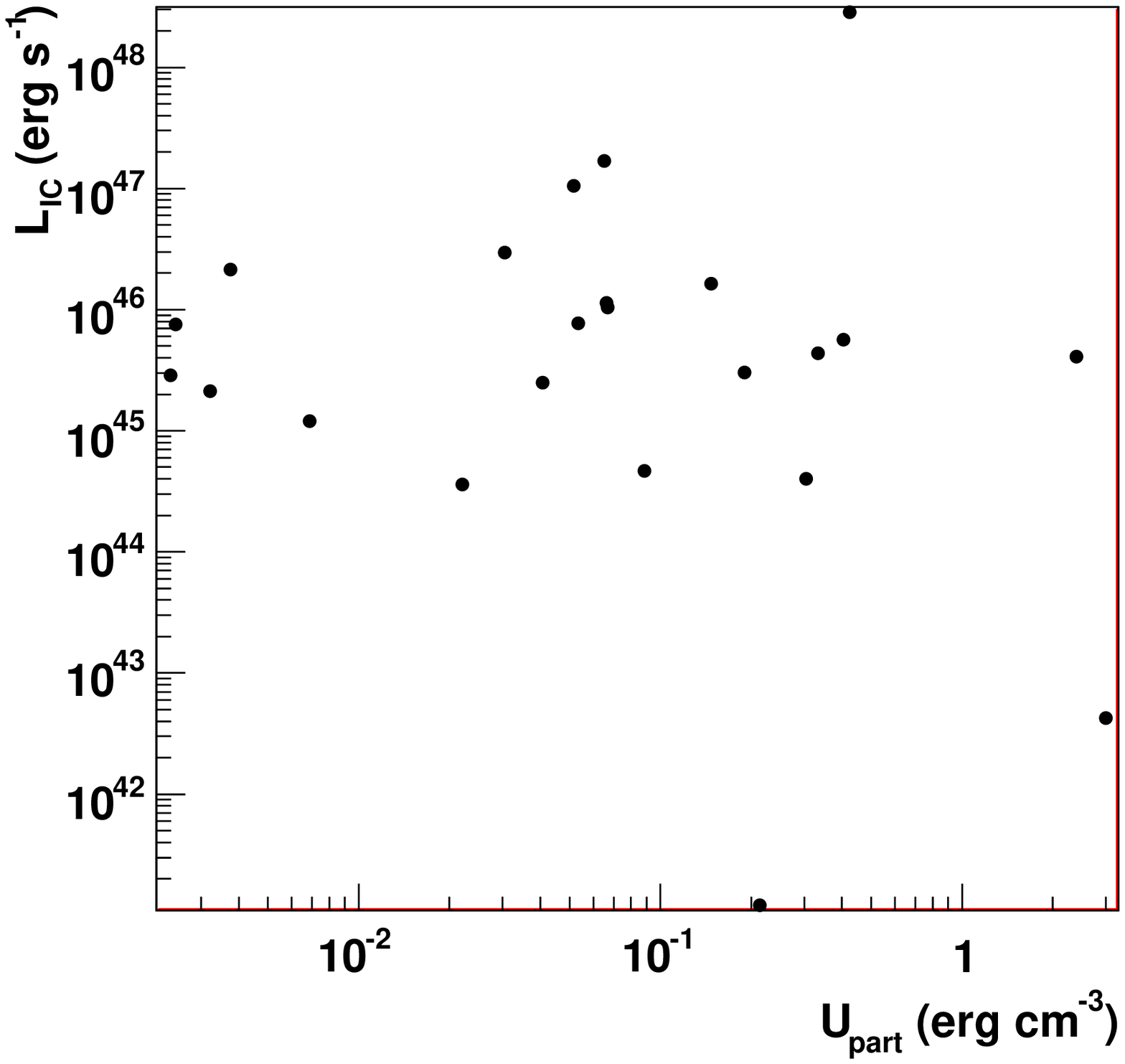,width=0.45\textwidth}
}
\vspace*{8pt}
\caption{{\it Left}: Inverse Compton luminosity $L_\mathrm{IC}$ versus the peak frequency of the inverse Compton component $\nu_c^\mathrm{peak}$.
{\it Right}: Inverse Compton luminosity $L_\mathrm{IC}$ versus particle energy density $U_\mathrm{part}$.
\label{fig-UpartLIC_NuICPeakLIC}}
\end{figure}

In Fig.~\ref{fig-UpartLIC_NuICPeakLIC}, we show the inverse Compton luminosity $L_\mathrm{IC}$ in function of the peak frequency of the inverse Compton component $\nu_c^\mathrm{peak}$. Contrary to what is expected from the blazar sequence\cite{1998MNRAS.299..433F}, no correlation is found ($\rho=-0.20$). This lack of correlation, at high frequencies, could be due to the instrumental sensitivity limit.
$L_\mathrm{IC}$ is also expected to be correlated with the particle energy density $U_\mathrm{part}$, but no evident correlation appeared in our study ($\rho=0.01$). One might then think the blazar sequence could be questionable as such, as was already suggested in Refs.~\refcite{2008A+A...488..867N,2007Ap+SS.309...63P}.

\section{Conclusion}

In a future paper, we will present the SEDs for the complete set of currently detected AGNs in the VHE domain, all consistently interpreted in the same SSC framework. We will discuss further the parameter study and investigate its implication on the blazar sequence and on individual properties of these objects.

\section*{Acknowledgments}

{\scriptsize
      J.-P.~L. acknowledges receipt of a fellowship funded by the European Commission's Framework Programme 6, through the Marie Curie EST project MEST-CT-2005-021074. J.-P.~L. would like to warmly thank all the colleagues from the Astronomy group of the Department of Physics and Astronomy of Durham University, U.K. for their welcome during his visit. We would like to thank a lot of people from the VHE astronomy community, too numerous to be quoted here, who kindly shared their VHE data.

}


\end{document}